\documentclass[aps,prl,superscriptaddress,reprint,amsmath,amssymb,letterpaper,showpacs]{revtex4-1}

\usepackage{txfonts}
\usepackage{graphicx}
\usepackage{subfigure}
\usepackage{color}

\newcommand{\tr}{\textrm} 
\newcommand{\mb}{\mathbf} 

\begin{document}

\allowdisplaybreaks


\title{Femtosecond 240-keV electron pulses from direct laser acceleration in a low-density gas}

\author{Vincent Marceau}
\affiliation{Centre d'Optique, Photonique et Laser, Universit\'e Laval, Qu\'ebec (QC) G1V 0A6, Canada}
\author{Charles Varin}
\affiliation{Center for Research in Photonics, University of Ottawa, Ottawa (ON) K1N 6N5, Canada}
\author{Thomas Brabec}
\affiliation{Center for Research in Photonics, University of Ottawa, Ottawa (ON) K1N 6N5, Canada}
\author{Michel Pich\'e}
\affiliation{Centre d'Optique, Photonique et Laser, Universit\'e Laval, Qu\'ebec (QC) G1V 0A6, Canada}

\date{\today}

\begin{abstract}
We propose a simple laser-driven electron acceleration scheme based on tightly focused radially polarized laser pulses for the production of femtosecond electron bunches with energies in the few-hundreds-of-keV range. In this method, the electrons are accelerated forward in the focal volume by the longitudinal electric field component of the laser pulse. Three-dimensional test-particle and particle-in-cell simulations reveal the feasibility of generating well-collimated electron bunches with an energy spread of 5\% and a temporal duration of the order of 1 fs. These results offer a route towards unprecedented time resolution in ultrafast electron diffraction experiments. 
\end{abstract}


\maketitle

The development of high-power laser facilities all around the world has paved the way to the design of a new generation of laser-based electron accelerators. Recent experimental successes have shown that electrons may be accelerated to hundreds-of-MeV energies from high-intensity laser-plasma interactions~\cite{banerjee12_physplasmas,wang12_aipconf,banerjee13_prstab}. Laser-driven electron accelerators are thus expected to offer a robust, compact, and low-cost alternative to conventional radio-frequency (rf) accelerators~\cite{malka06_ptrsa}. 

While most studies have been concerned with the laser acceleration of electron bunches to energies ranging from several MeV to the GeV level~\cite{esarey09_rmp}, comparatively little work has been done at lower energies (e.g.,~\cite{tokita09_apl,uhliga11_lpb,payeur12_apl,he13_njp,breuer13_prl}). 

In fact, due to their large scattering cross section in comparison to x-rays, subrelativistic electrons find important applications in atomic and molecular imaging experiments~\cite{sciaini11_rpp}. In the last few years, electrons at subrelativistic energies have been successfully used in time-resolved ultrafast electron diffraction (UED) experiments to study dynamical processes on the subpicosecond time scale~\cite{harb08_prl,sciaini09_nature,gao13_nature}. In the latter experiments, the electrons are generated from the illumination of a photocathode by a femtosecond laser pulse and are subsequently accelerated in a static electric field. Using this method, electron bunches with a duration between 200 and 350~fs and energy in the 50--100~keV range can be produced~\cite{sciaini11_rpp}. In addition, using state-of-the-art rf cavities to invert the linear velocity chirp, the electron bunches can be compressed down to about 70~fs at the sample~\cite{vanoudheusden10_prl}, while the timing jitter between the laser and the rf electronics can be reduced to 30~fs with the time stamping method~\cite{gao13_apl}. Bunches of shorter durations ($\sim$10~fs) have been predicted by replacing the static accelerator with a rf gun that accelerates the electrons at energies of a few MeV~\cite{han11_prstab}. However, due to the reduced scattering cross section of relativistic electrons and other practical considerations, the 100--300~keV energy window is generally preferred for UED~\cite{vanoudheusden07_jap}.

Recently, laser-driven electron acceleration has been proposed as an alternative to static accelerator technology for UED experiments~\cite{tokita09_apl,tokita10_prl,he13_apl}. In principle, laser acceleration has several advantages~\cite{he13_apl}: (i) the short wavelength of the accelerating field may lead to electron bunches with duration of the order of 10~fs or less; (ii) there is an intrinsic synchronization between the electron probe and the laser pump; (iii) using a gas medium, the electron source is self-regenerating and can thus be used for experiments at high repetition rates. In~\cite{tokita10_prl}, 350-keV electron bunches were produced from a high-intensity laser-solid interaction and compressed down to 500 fs, while in~\cite{he13_apl}, 100-keV bunches with a duration possibly under 100 fs, although not measured, were generated with a laser-wakefield accelerator. Subfemtosecond electron pulses are predicted in plasmas with ramp-up density profiles, but at relativistic energies~\cite{li13_prl}.

In this Letter, we propose a simple direct acceleration scheme based on the use of tightly focused radially polarized laser pulses for the generation of electron bunches with unprecedentedly short duration in an energy range appropriate for UED applications. This method takes advantage of the strong longitudinal electric field at the beam center to accelerate the electrons from the focal region along the optical axis~\cite{varin13_applsci}. We demonstrate the feasibility of generating 240-keV, one-femtosecond electron pulses when the laser pulse is tightly focused in a low-density gas. The acceleration mechanism is first analyzed using a three-dimensional test-particle approach. We then investigate the limits of validity of these results using three-dimensional particle-in-cell (3DPIC) simulations with full ionization dynamics. We finally discuss how the proposed acceleration scheme could find applications in time-resolved UED experiments.

Ultrashort and tightly focused laser pulses must be modeled as exact solutions to Maxwell's equations. We consider the lowest-order radially polarized laser field, namely a TM$_{01}$ pulse, for which an exact closed-form solution is known~\cite{april10_intech,marceau12_optlett}. In vacuum, the nonzero field components of a TM$_{01}$ pulse traveling in the forward $z$ direction with its beam waist plane located at $z=0$ are given, in cylindrical coordinates ($r,\theta,z$), by the following expressions:
\begin{align}
&\!\! E_r (\mb{r},t) = \Re\bigg\{ \frac{3 E_0 \sin 2\tilde{\Theta}}{2\tilde{R}} \bigg( \frac{G_-^{(0)}}{\tilde{R}^2} \!+\! \frac{G_+^{(1)}}{c\tilde{R}} \!+\! \frac{G_-^{(2)}}{3c^2}\bigg) \bigg\} \ ,  \label{eq:npTM01Er}\\
&\!\! E_z (\mb{r},t) =  \Re\bigg\{ \frac{E_0}{\tilde{R}} \bigg[  \frac{(3\cos^2\tilde{\Theta}\!-\!1)}{\tilde{R}}  \bigg( \frac{G_-^{(0)}}{\tilde{R}} \!+\! \frac{G_+^{(1)}}{c} \bigg) \!-\! \frac{\sin^2\tilde{\Theta}}{c^2} G_-^{(2)} \bigg] \bigg\}  \ , \label{eq:npTM01Ez} \\ 
&\!\! H_\phi (\mb{r},t) = \Re\bigg\{ \frac{E_0 \sin \tilde{\Theta}}{\eta_0 \tilde{R}} \bigg( \frac{G_-^{(1)}}{c\tilde{R}} \!+\! \frac{G_+^{(2)}}{c^2}\bigg) \bigg\} \ , \label{eq:npTM01Hphi}
\end{align}
where $\Re\{\cdots\}$ denotes the real part, $E_0$ is a constant amplitude, $c$ is the speed of light in vacuum, $\tilde{R}=[r^2 + (z+ja)^2]^{1/2}$, $\cos \tilde{\Theta} = (z+ja)/\tilde{R} $, and $G^{(n)}_\pm = \partial^n_t [f(\tilde{t}_-)\pm f(\tilde{t}_+)]$ with $\tilde{t}_\pm = t \pm \tilde{R}/c + ja/c$. The confocal parameter $a$ can be used to characterize the degree of paraxiality of the beam since it is related to the Rayleigh range $z_R$ at wavelength $\lambda_0$ by $k_0 z_R = \sqrt{1+(k_0 a)^2} - 1$~\cite{rodriguez-morales04_optlett}. The function $f(t)$ is the inverse Fourier transform of the frequency spectrum of the pulse, which we assume to be Poisson-like~\cite{caron99_jmodoptic}:
\begin{align}
F(\omega) = 2\pi e^{-j\phi_0} \left( \frac{s}{\omega_0}\right)^{s+1} \frac{\omega^s e^{-s\omega/\omega_0}}{\Gamma(s+1)}\ H(\omega) \ ,
\end{align}
where $s$ is a positive parameter related to the pulse duration, $\phi_0$ is the constant pulse phase, $\omega_0=ck_0$ is the frequency of maximum amplitude, $\Gamma(x)$ is the gamma function, and $H(x)$ is the Heaviside step function. The fields given by Eqs.~\eqref{eq:npTM01Er}--\eqref{eq:npTM01Hphi} may be produced by focusing a collimated radially polarized input beam with a parabolic mirror of large aperture~\cite{april10_optexpress}. 

Conceptually, our accelerator design simply consists of an ultrashort TM$_{01}$ pulse that is strongly focused in a low-density gas target of uniform density $n_0$. This configuration is very similar to the experimental setup recently used by Payeur \emph{et al.}~\cite{payeur12_apl}. To simulate the laser-driven electron acceleration, we perform three-dimensional simulations using the EPOCH PIC code~\cite{brady11_ppcf}. The fields given in Eqs.~\eqref{eq:npTM01Er}--\eqref{eq:npTM01Hphi} are implemented in the code using the scattered-field formulation~\cite{taflove05_book}.

\begin{figure}[!t]
\centering
\includegraphics[width=0.48\columnwidth]{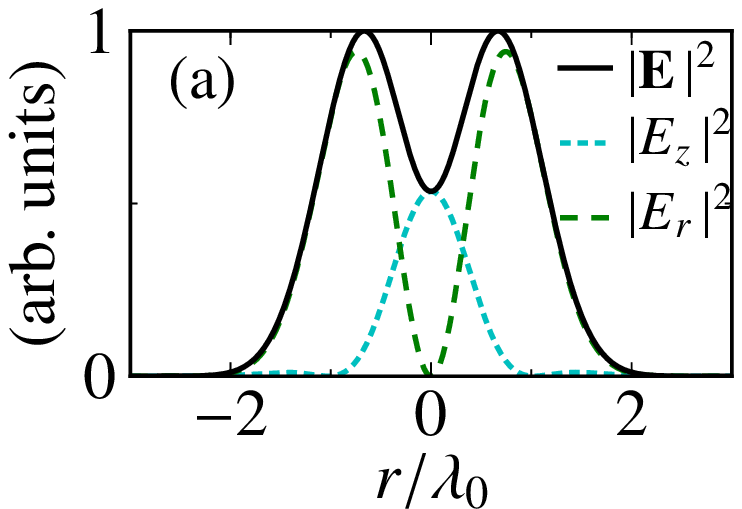} \ \
\includegraphics[width=0.48\columnwidth]{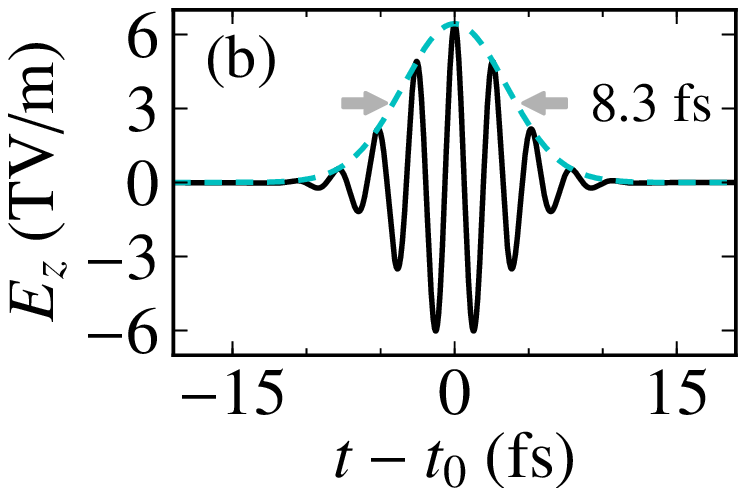}
\caption{(Color online) Electric field of a TM$_{01}$ pulse with $k_0 a = 20$, $s=70$, $P=300$ GW, and $\lambda_0 = 800$ nm. (a) Electric energy density at the beam waist. (b) Temporal variation of $E_z$ at the origin. \label{fig:TM01}}
\end{figure}

We consider a TM$_{01}$ pulse characterized by $k_0a = 20$ and $s = 70$ with an average power of $P=300$~GW and a dominant wavelength of $\lambda_0 = 800$~nm. This gives a pulse with a full width at half maximum (FWHM) duration of 8.3~fs. The center of the pulse is set to reach the beam waist plane at $t_0=80$~fs. The spatiotemporal properties of the pulse in vacuum at the beam waist are illustrated in Fig.~\ref{fig:TM01}. The chosen laser parameters correspond to a regime accessible by current millijoule lasers that can operate at kHz repetition rate with carrier-envelope phase stabilization~\cite{mashiko07_apl,chen11_laserphys}. 

In analogy to the standard normalized vector potential parameter $a_0=e|E|/m_e c \omega_0$~\cite{hartemann01}, it is useful to introduce a normalized longitudinal field parameter $a_z = e |E_z|_\tr{peak}/m_e c \omega_0$~\cite{varin13_applsci}. At $a_z\gtrsim1$, the motion of a free electron in the longitudinal electric field becomes relativistic. Consequently, \textit{subcycle acceleration}, i.e., the process in which the electron is accelerated by staying in phase with the laser field over a certain distance, starts to take place~\cite{varin05_pre}. On the one hand, subcycle acceleration induces a strong longitudinal compression over a cloud of electrons, which promotes the formation of ultrashort electron bunches~\cite{varin06_pre,karmakar07_lpb}. On the other hand, if the value of $a_z$ is too high, the electrons will acquire an energy that will be too great for any use in electron diffraction experiments. For the chosen laser pulse parameters, we have $a_z \approx 1.6$, a good tradeoff between the longitudinal compression induced by subcycle acceleration and the final kinetic energy of the electrons. 

\begin{figure}[!t]
\centering
\includegraphics[width=\columnwidth]{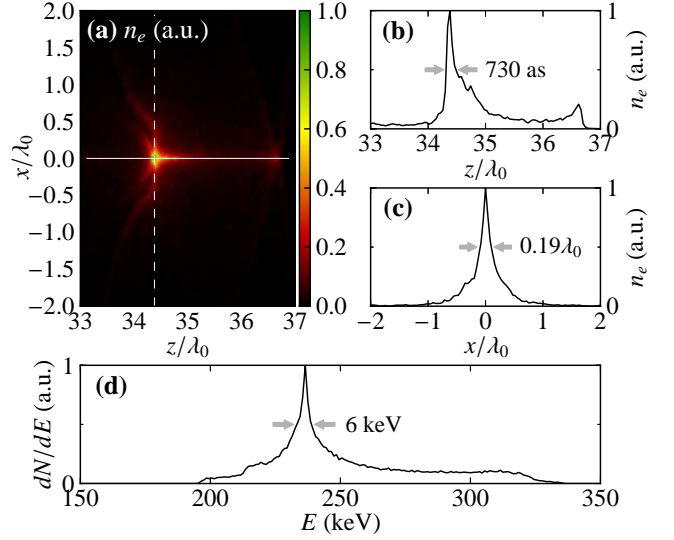}
\caption{(Color online) Characterization of an ultrashort electron bunch produced after the interaction of a $k_0a = 20$, $s=70$, $P=300$~GW, $\phi_0 = \pi$ laser pulse with a target of free and noninteracting electrons. (a) Electron number density $n_e$ in the $(x,z)$ plane. The solid and dashed white lines indicate the locations of the $z$ and $x$ cuts shown in (b) and (c), respectively. The arrows in (b) and (c) indicate the full width at half maximum duration and width. (d) Kinetic energy distribution. The snapshot is taken at $t-t_0=120$~fs. Only the electrons with $\Delta W > 50$~keV located in the ($x$,$z$) region shown in (a) and within a slice of thickness $\lambda_0$ centered at $y=0$ are considered to obtain the distribution functions. The simulation is performed using $10^9$ pseudoparticles randomly distributed in the region $x,y \in [-5\lambda_0,5\lambda_0]$, $z \in [-14\lambda_0,50\lambda_0]$.  \label{fig:number_density_tracer}}
\end{figure}

To investigate the acceleration dynamics, we adopt in the first place a three-dimensional test-particle approach in which all electrons are initially assumed to be free and space-charge effects are neglected. The electrons are initially distributed randomly in space according to a uniform distribution. As the laser pulse approaches the focal region, two main electron jets are formed: an annular electron jet is accelerated away from the optical axis under the influence of the radial electric field component, and a well-collimated electron bunch is accelerated in the forward $z$ direction by the longitudinal electric field component. Figure \ref{fig:number_density_tracer} illustrates the main properties of this on-axis electron bunch. At the instant the snapshots shown in Fig.~\ref{fig:number_density_tracer} are taken, the interaction of the electron bunch with the laser pulse is already terminated. The divergence of the bunch is estimated to be 6~mrad, while its duration, given by the longitudinal extent of the bunch divided by the average velocity of the electrons, is of the order of 730~as. The energy distribution of the electron bunch, shown in Fig.~\ref{fig:number_density_tracer}(d), displays a well-defined maximum at $E = 237$~keV with a small absolute energy spread of $\Delta E = 6$~keV. Accelerating subfemtosecond pulses with radially polarized laser pulses using an infinite target is noteworthy, as previous techniques had to rely on nanometric targets and ultrarelativistic laser intensities ($a_z^2 \gg  1$)~\cite{varin06_pre,karmakar07_lpb}.

\begin{figure}[!t]
\centering
\includegraphics[width=\columnwidth]{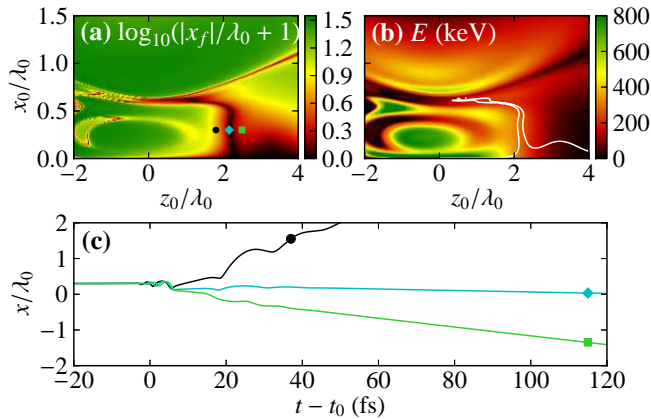}
\caption{(Color online) (a) Final $x$ coordinate [$\tr{log}_{10}(|x_f|/\lambda_0 + 1)$] and (b) final kinetic energy at time $t-t_0=120$ fs for a free electron initially at rest at the position $(x_0,z_0)$ in the $(x,z)$ plane. The three markers in (a) correspond to three initial conditions for which the trajectory is illustrated in (c). The white curve in (b) corresponds to the $|x_f|=0.5\lambda_0$ contour. All laser parameters are identical to those used in Fig.~\ref{fig:number_density_tracer}. \label{fig:acceleration_z0r0}}
\end{figure}

To get a better understanding of the formation of the ultrashort electron pulse reported in Fig.~\ref{fig:number_density_tracer}, it is instructive to identify the origin of the electrons of which it is made. Figure \ref{fig:acceleration_z0r0}(a) maps the initial coordinates of a free electron in the $(x,z)$ plane to its final transverse coordinate. The most remarkable feature is the presence of a thin vertical band at $z_0\approx 2.15\lambda_0$ extending from $x_0=0$ to $x_0\approx0.5\lambda_0$ that corresponds to electrons that remain within a distance of $\lambda_0/2$ from the optical axis. Moreover, as shown in Fig.~\ref{fig:acceleration_z0r0}(b), this set of initial conditions is correlated with a region where the final kinetic energy of the electrons is extremely similar. Therefore, the formation of an ultrashort electron pulse originates from the acceleration of a thin disk of electrons located in a very restricted region of the infinite gas target. Electrons outside this thin disk region are either deflected away from the optical axis [see Fig.~\ref{fig:acceleration_z0r0}(c)] or gain little energy from the laser field. We emphasize that the acceleration process is sensitive to the carrier-envelope phase, a clear signature of direct acceleration that distinguishes our scheme from ponderomotive acceleration, which is a process independent of the laser pulse phase~\cite{stupakov01_prl}. Here, substantial energy gains are possible because an asymmetry between consecutive positive and negative half field cycles is introduced by nonlinear relativistic effects, the ultrashort (few-cycle) pulse duration, and the strong field divergence ($z_R\sim 3\lambda_0$).

\begin{figure}[!t]
\centering
\includegraphics[width=\columnwidth]{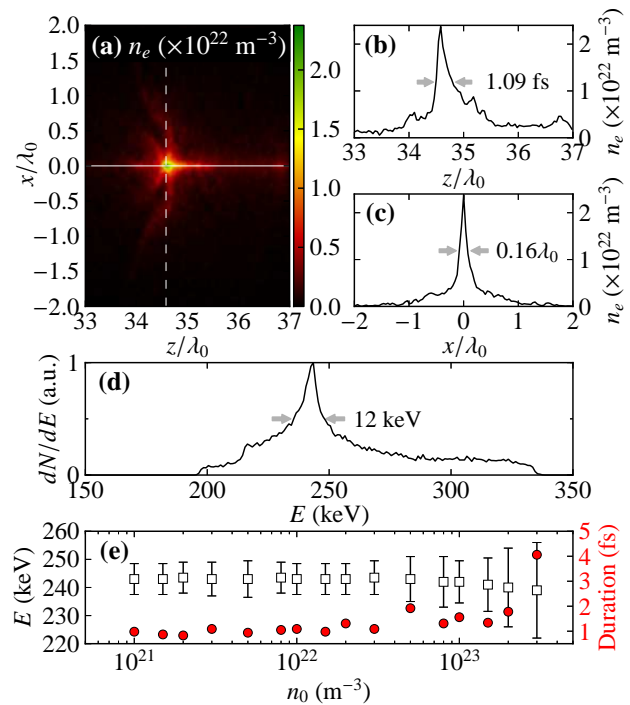}
\caption{(Color online) (a)--(d) Three-dimensional particle-in-cell simulation corresponding to the case shown in Fig.~\ref{fig:number_density_tracer}. The target consists of neutral hydrogen with a uniform density of $n_0 = 3 \times 10^{22}$~m$^{-3}$. (e) Variation of the peak energy and energy spread (square markers with error bars, left scale), and duration (circle markers, right scale) of the electron pulse as a function of the initial atomic density $n_0$. In each case, the simulation is performed using $5 \times 10^8$ hydrogen pseudoparticles randomly distributed on a $200 \times 200 \times 1000$ grid with $N_\lambda=20$ points per wavelength resolution, corresponding to the region $x,y \in [-5\lambda_0,5\lambda_0]$, $z \in [-10\lambda_0,40\lambda_0]$. The statistics are obtained at $t-t_0 = 120$~fs. Increasing the grid resolution to $N_\lambda=30$ does not significantly alter the results.\label{fig:number_density_hydrogen}}
\end{figure}

Having studied the acceleration mechanism in the single particle limit, we now proceed with a much more realistic 3DPIC approach. We assume that the initial target consists of a neutral hydrogen gas of uniform atomic density $n_0$ at room temperature. Multiphoton, tunnel, and barrier-suppression ionization are taken into account~\cite{lawrence_phdthesis}. Figure \ref{fig:number_density_hydrogen}(a)--(d) show the results of a simulation performed at an initial density of $n_0=3\times10^{22}$~m$^{-3}$ with the same laser parameters as in Fig.~\ref{fig:number_density_tracer}. The electron pulse produced from the hydrogen target possesses features very similar to that reported in Fig.~\ref{fig:number_density_tracer}. Its duration is slightly above one femtosecond, with a peak areal density and a total charge of $2\times10^{-3}$ C$\cdot$m$^{-2}$ and 1.1~fC, respectively. The charge was obtained by counting up the electrons located within a cylinder of radius $\lambda_0$ extending from $z=33\lambda_0$ to $z=37\lambda_0$ [see Fig.~\ref{fig:number_density_hydrogen}(a)]. We calculate the fraction of the electrons within the FWHM of the longitudinal and radial density distribution to be 2.4\%. We emphasize that due to the ionization dynamics, other elements than hydrogen might not be used to generate monoenergetic electron pulses. In fact, we have observed that 3DPIC simulations with helium yield an energy distribution with two distinct peaks (not shown).

In Fig.~\ref{fig:number_density_hydrogen}(e), we illustrate the variation of the main features of the electron pulse as a function of the initial density $n_0$. Well-collimated, monoenergetic one-femtosecond pulses are observed up to densities of about $n_0=3\times10^{22}$~m$^{-3}$. Up to this density, the bunch charge increases linearly. As $n_0$ is raised above $3\times10^{22}$~m$^{-3}$, the electron pulse duration and its energy spread increase under the influence of electrostatic repulsion. At initial atomic densities of $n_0=3\times10^{23}$~m$^{-3}$ and above, the electron pulse is rapidly broadened by space-charge forces. 

The possibility of generating femtosecond monoenergetic electron bunches suggests that the proposed 
 acceleration scheme could offer an interesting avenue towards unprecedented time resolution in UED 
 experiments. In this perspective, the \textit{transverse normalized emittance} and the \textit{transverse 
 coherence length} are important parameters. The normalized emittance, which estimates the volume occupied 
 by the electron beam in phase space, is given, in the $x$ direction, by $\epsilon_{n,x} = (1/m_e c) 
 \sqrt{\langle x^2 \rangle \langle p_x^2 \rangle - \langle xp_x \rangle^2 }$, where $\langle \cdots \rangle$ 
 denotes an average over the electrons in the bunch~[18]. With the electrons from the same region 
 that was used to calculate the bunch charge in  Fig.~5(a) we get
 $\epsilon_{n,x}=\epsilon_{n,y}=3.6\times10^{-3}$~mm$\cdot$mrad, which compares favorably to 
 state-of-the-art UED setups ($\epsilon_{n,x} \approx 2 \times 10^{-2}$ mm$\cdot$mrad~[18,20]). The 
 transverse coherence length is calculated with $L_c = \lambda / 2\pi \sigma_\theta$, where $\lambda$ is 
 the de Broglie wavelength and $\sigma_\theta$ is the root-mean-square angular spread~[18]. Here we obtain $L_c  =  0.03$ nm, which is too small for UED. Nevertheless, we estimate that 
 filtering the electron pulses with a pinhole would remove the most divergent electrons and increase the 
 transverse coherence length beyond 1~nm. With the 
 remaining electrons per pulse ($\sim 10^2-10^3$), an electron flux sufficient for 
 time-resolved crystallography experiments would be possible at a
 kHz repetition rate (see, e.g., \cite{BaumScience2007}). Finally, we recall that for the proposed scheme  
 ---which is a subcycle laser process--- the energy spread is 
 intrinsically low (about 5\%). Energy filtering as done in \cite{tokita10_prl} might be beneficial, but is probably not necessary.
 
Still in connection with UED, we stress that free-space propagation is needed for the electron pulse to reach the sample, which causes spatiotemporal broadening. Nevertheless, the use of a strongly focused, rapidly diverging, laser pulse ($z_R \sim 3 \lambda_0$) could allow having the sample very close to the focus. With the available computational resources, we were able to simulate the propagation of the electron pulse for $1.1$ ps. Whereas the radius of the electron pulse did not change significantly, the pulse duration, after an initial transient behavior, increased linearly at the rate of $0.027$~fs/$\mu$m. This asymptotic linear behavior is in agreement with existing theoretical models~\cite{siwick02_jap}. It thus appears that pulse compression will be needed to keep the duration at the sample below 30~fs for a focus-sample distance larger than 1~mm. For state-of-the-art compression techniques see~\cite{vanoudheusden10_prl,tokita10_prl}.
 
Finally, we emphasize that besides UED, the proposed acceleration scheme might also be of interest for electron injection into x-ray free electron lasers and staged (channel-guided) laser wakefield accelerators, as well as for the development of tabletop radiation sources. For those applications, relativistic energies are needed. Preliminary results for the actual acceleration scheme show that it would be possible to double the electron energy only by tuning the laser power, while preserving a good beam quality ($\Delta E /E \sim 10\%$ or less). The laser pulse parameters were not submitted to an intensive optimization process. Reaching the few-MeV range is the object of ongoing research.  

\begin{acknowledgments}
This research was supported by the Natural Sciences and Engineering Research Council of Canada (NSERC). The authors gratefully acknowledge Calcul Qu\'ebec -- Universit\'e Laval and Compute Canada for computational resources and support, as well as the EPOCH development team. 
\end{acknowledgments} 


\end{document}